\documentclass[journal]{IEEEtran}

\usepackage{cite, setspace,graphics, subfigure, amsmath,amssymb,amsfonts,stfloats, url}
\usepackage{lipsum}
\usepackage{mathtools}
\usepackage{cuted}\stripsep -3pt plus 3pt minus 2pt
\usepackage[below]{placeins}
\usepackage{booktabs}
\usepackage{multirow}
\usepackage{graphicx}
\usepackage[normalem]{ulem}
\useunder{\uline}{\ul}{}
\usepackage{pifont}
\usepackage{enumerate} 

\DeclareGraphicsExtensions{.pdf,.jpeg,.png}
\ifCLASSINFOpdf
\else
\fi

\ifCLASSOPTIONcompsoc
 \usepackage[caption=false,font=normalsize,labelfont=sf,textfont=sf]{subfig}
\else
 \usepackage[caption=false,font=footnotesize]{subfig}
\fi

\begin{document}
\title{High Capacity Lossless Data Hiding in JPEG Bitstream Based on General VLC Mapping}

\author{Yang Du,~Zhaoxia~Yin,~\IEEEmembership{Member,~IEEE},~and~Xinpeng Zhang,~\IEEEmembership{Member, IEEE}
\thanks{This research work is partly supported by National Natural Science Foundation of China (61872003, U1636206,61860206004).}
\thanks{Yang Du and Zhaoxia Yin are with the school of Computer Science and Technology, Anhui University, e-mail: yinzhaoxia@ahu.edu.cn.}
\thanks{Xinpeng Zhang was with the school of Computer Science, Fudan University, e-mail: zhangxinpeng@fudan.edu.cn.}}

\markboth{IEEE Transactions on Dependable and Secure Computing}
{Du \MakeLowercase{\textit{et al.}}: High Capacity Lossless Data Hiding in JPEG Bitstream Based on General VLC Mapping}

\maketitle

\begin{abstract}
JPEG is the most popular image format, which is widely used in our daily life. Therefore, reversible data hiding (RDH) for JPEG images is important. Most of the RDH schemes for JPEG images will cause significant distortions and large file size increments in the marked JPEG image. As a special case of RDH, the lossless data hiding (LDH) technique can keep the visual quality of the marked images no degradation. In this paper, a novel high capacity LDH scheme is proposed. In the JPEG bitstream, not all the variable length codes (VLC) are used to encode image data. By constructing the mapping between the used and unused VLCs, the secret data can be embedded by replacing the used VLC with the unused VLC. Different from the previous schemes, our mapping strategy allows the lengths of unused and used VLCs in a mapping set to be unequal. We present some basic insights into the construction of the mapping relationship. Experimental results show that most of the JPEG images using the proposed scheme obtain smaller file size increments than previous RDH schemes. Furthermore, the proposed scheme can obtain high embedding capacity while keeping the marked JPEG image with no distortion.
\end{abstract}

\IEEEpeerreviewmaketitle

\section{Introduction}
\IEEEPARstart{R}{DH} can embed additional data into cover media imperceptibly while the cover media can be reconstructed losslessly during the data extraction process. The reversibility feature is desirable for many applications, such as medical or military image processing and law forensics. Recently, RDH is extended to many other fields innovatively\cite{IEEEexample:hou2019emerging}, such as reversible steganography\cite{IEEEexample:hong2015reversible}\cite{IEEEexample:zhang2015reversible}, reversible adversarial example\cite{liu2018reversible} and reversible data hiding in encrypted domain\cite{qian2016separable}\cite{zheng2019lossless}.
\par Many RDH schemes for uncompressed images have been proposed and can be classified into three categories based on different technologies: lossless compression \cite{fridrich2002lossless}\cite{celik2005lossless}, difference expansion (DE)\cite{IEEEexample:tian2003reversible}\cite{dragoi2014local}, and histogram shifting (HS)\cite{IEEEexample:ni2006reversible,zhang2012reversible,zhang2013recursive,jia2019reversible}. But these RDH schemes are designed for uncompressed images. However, JPEG is the most popular image format which is applied to digital cameras and the Internet widely. Therefore, developing the RDH scheme for JPEG images is more important.
\par The existing RDH schemes for JPEG images can be classified into two categories according to the modifying domain. The first category of schemes [15]-[24] is based on modifying the Discrete Cosine Transform (DCT) coefficients domain. We refer to this category of schemes as DCT-based schemes. The second category of schemes [24]-[27] is based on modifying the VLC domain. We refer to this category of schemes as VLC-based schemes.
\par Some early works \cite{IEEEexample:fridrich2002lossless, IEEEexample:wang2013high} of DCT-based schemes modified the quantization table in the file header to embed data. In \cite{IEEEexample:fridrich2002lossless}, Fridrich ${ et~al. }$ firstly proposed an RDH scheme that modified the quantization table factors to embed one bit per DCT coefficient. The basic idea is to divide some quantized table factors by 2 and multiply the quantized DCT coefficients of their corresponding positions by 2 to embed the information into the least significant bits (LSB) of DCT coefficients. Wang ${ et~al. }$ \cite{IEEEexample:wang2013high} designed a new embedding order to embed information in a position preferentially where distortion is small. In \cite{IEEEexample:fridrich2004lossless}, Fridrich ${ et~al. }$ proposed a scheme based on lossless compression of quantized DCT coefficients. This scheme creates a vacated room by compressing the generalized least significant bits of non-zero alternating current (AC) coefficients. However, the LSB distribution of the non-zero AC is still balanced relatively. Therefore, the embedding capacity is rather limited.
\par Statistically, the quantized AC coefficient distribution is a zero-centered Laplacian-like distribution, so the literature \cite{IEEEexample:xuan2007reversible, IEEEexample:li2010reversible} migrated the HS technique from uncompressed images to JPEG images. \cite{IEEEexample:chen2014recursive} constructed a ternary optimal RDH code based on AC coefficients with values of 0, 1, and -1. But changing the AC coefficients with a value of 0 leads to file size expansion inevitably. To further improve the performance of the above HS-based methods for JPEG images, Huang ${ et~al. }$ \cite{IEEEexample:huang2016reversible} proposed a novel HS-based scheme. It keeps the zero AC coefficient unchanged and sets the AC coefficients with values of 1 and -1 as the peak points of the histogram. Since the variation of all non-zero AC coefficients is at most 1, the visual quality is guaranteed. Besides, to further reduce the distortion, Huang ${ et~al. }$ designed a block selection strategy based on the number of zero coefficients in each 8 ${\times}$ 8 block. However, the number of zero coefficients in a block does not imply the degree of distortion of the block. In \cite{IEEEexample:qian2017reversible}, Qian ${ et~al. }$ analyzed the features of JPEG and then proposed an ordered embedding strategy in several rounds to reduce the file size increments. In \cite{IEEEexample:wedaj2017improved, IEEEexample:hou2018reversible}, to optimize the block selection strategy, the influence of the quantization step in DCT coefficients are considered. These schemes can keep the good visual quality and also achieve a smaller file size expansion.
\par The previous VLC-based schemes embed secret data based on VLC mapping, which can preserve the visual quality of the marked image unchanged. In \cite{IEEEexample:qian2012lossless}, Qian and Zhang proposed a scheme to embed data into the JPEG bitstream by replacing the used VLCs with unused VLCs. The visual quality and file size are both preserved. Hu ${ et~al. }$  and Qiu ${ et~al. }$ \cite{IEEEexample:hu2013improved} \cite{IEEEexample:qiu2018lossless}explored the redundancy of VLCs in JPEG bitstream, and further improved the embedding capacity by VLC frequency ordering and mapping optimizing. However, the embedding capacity is related to the frequency of used VLCs, which is rather limited. Unfortunately, at present, the developments of VLC-based RDH schemes are far from enough than DCT-based schemes.
\par For the most RDH schemes in JPEG images, the basic evaluation metrics are embedding capacity, visual quality, and file size preservation. Generally, there exists a trade-off between visual quality and embedding capacity. Similarly, there also exists a trade-off between file size preservation and embedding capacity. For the DCT-based RDH schemes, the optimization for all the three evaluation metrics is required. However, for the DCT-based RDH schemes, file size preservation and visual quality are difficult to be the optimal simultaneously. While for the VLC-based schemes, only the optimization for embedding capacity and file size preservation is required. The comparison of optimization objectives for the two kinds of schemes is depicted in Fig. 1. 
\begin{figure}[ht]
\centering
\includegraphics[width=3.3 in]{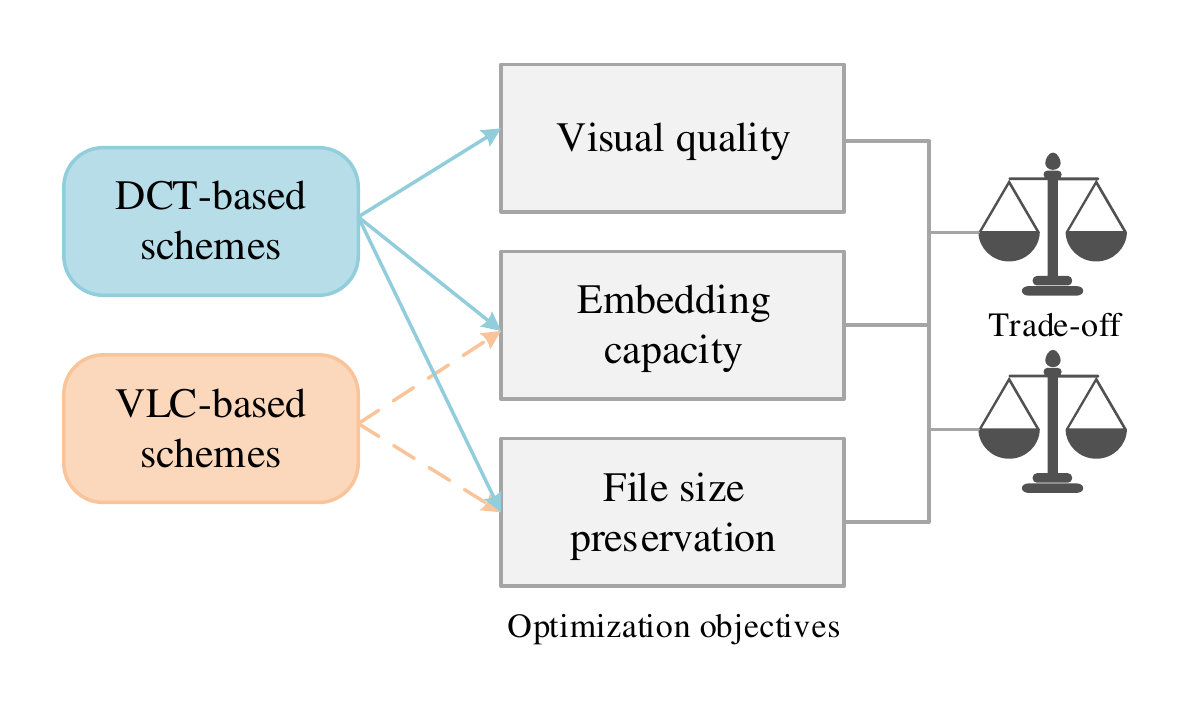}
\caption{The comparison of optimization objectives between DCT-based scheme and VLC-based scheme.}
\label{fig_trade_off}
\end{figure}
\par However, the previous VLC-based schemes are focusing on improving the embedding capacity while preserving the file size unchanged. Although the VLC-based schemes maintain the file size and visual quality unchanged, the embedding capacity is low indeed. Compared to this case, the high embedding capacity with slight file size expansion is more acceptable for most applications. 
\par In this paper, a general VLC mapping (GVM) strategy is adopted for LDH in JPEG bitstream. Using the GVM strategy, the lengths of VLCs may be unequal in one mapping set, so more used VLCs can be made use of to obtain high embedding capacity. Since the lengths of VLCs in one mapping set are unequal, the file size expansion is inevitable during the embedding process. The challenge is how to reduce the file size increments at a given payload as much as possible. We present some insights on how to construct the optimal GVM relationship while preserving the file size with an acceptable level. Experimental results demonstrate that via using the proposed GVM-based scheme, smaller file size increments can be obtained at different payloads and the image visual quality is also no degradation.
\par The remainder of this paper is organized as follows. In Section \uppercase\expandafter{\romannumeral2}, the JPEG bitstream structure and a basic embedding instance of VLC-based schemes are briefly introduced. Then the proposed GVM-based LDH scheme is proposed in Section \uppercase\expandafter{\romannumeral3}. The experimental results and analysis are presented in Section \uppercase\expandafter{\romannumeral4} to show the performance of the proposed scheme. Finally, this paper is concluded in Section \uppercase\expandafter{\romannumeral5}.

\section{Preliminaries}
\par In this section, the JPEG bitstream structure is first introduced briefly. Then, the basic embedding model of VLC-based schemes is described.
\subsection{JPEG Bitstream Structure}
\par The JPEG compression process consists of three phases: DCT, quantization and entropy encoding. In the entropy encoding phase, AC coefficients are pre-compressed by using the run-length encoding. The AC coefficients encoded by run-length encoding are represented as intermediate symbols of the form of ($run/size, value$). $run$ represents the number of successive zero AC coefficients before the current non-zero AC coefficient. The non-zero AC coefficient is encoded by Variable-Length Integer (VLI) coding. $size$ represents the length of the VLI code. $value$ represents the value of the next nonzero AC coefficient. Each $run/size$, which called Run/Size Value (RSV), is encoded with one VLC from the Huffman table for further compression. $value$ is encoded with VLI coding and the generated codes are called appended bits.
\par After the JPEG compression process. the entropy-coded data and other information are stored in the form of bitstream. The JPEG bitstream consists of the file header and entropy-coded data. JPEG file header includes some marker segments, such as the define Huffman table (DHT) marker segment. In the DHT segment, the Huffman table record all the RSVs and the lengths of corresponding VLCs. To decode the entropy-coded data with no error, each RSV is corresponded to a unique VLC. Fig. 2 shows the structure of the DHT segment, which contains the DHT header and the information to construct the Huffman table. The DHT header records the length of the segment and other information. ${L_{i}}$ is the number of VLCs of length ${i}$. All the VLCs can be generated according to the value from ${L_{1}}$ to ${L_{16}}$. ${V_{i,j}}$ is the RSV corresponds to the ${j}$-th VLC with the length of ${i}$. The entropy-coded data mainly consists of VLCs and appended bits. All the VLCs in the entropy-coded data can be found in the Huffman table.
\begin{figure*}[htbp]
\centering
\includegraphics[width = 5 in]{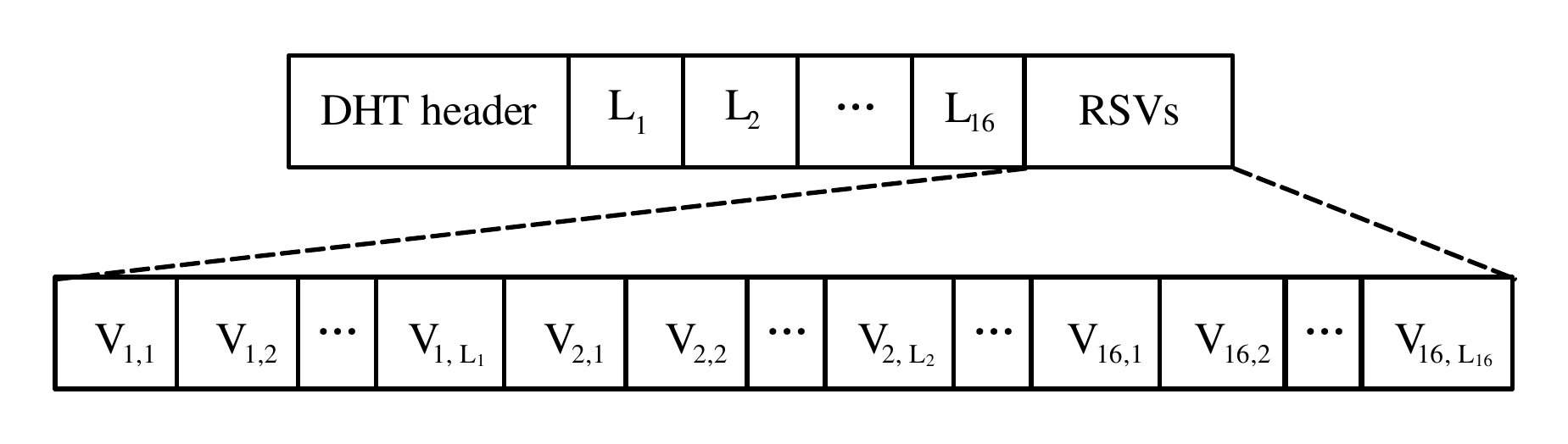} 
\caption{The structure of the DHT segment.}
\label{fig_dht}
\end{figure*}

\subsection{Basic embedding instance of VLC-based schemes}
\par In \cite{IEEEexample:qian2012lossless}, Qian and Zhang proposed a file size preserving LDH scheme by VLC mapping. For the used and unused VLCs, Qian and Zhang proposed a VLC mapping method, which maps the unused VLCs to the used VLC. To preserve the file size unchanged, the used and the mapped unused VLCs are the same lengths. While the visual quality is also unchanged by modifying the DHT segment in the file header. Fig. 2 illustrates an basic embedding instance using one mapping set, which the mapping set is  $\left\{VLC_1 \leftrightarrow\left \{ VLC_2,VLC_3,VLC_4 \right\} \right\}$. $VLC_1$ is a used VLC and the other VLC are unused VLCs. $VLC_1,VLC_2,VLC_3$ and $VLC_4$ represent the binary data "00", "01", "10" and "11" respectively. As Fig. 2 shows, When data "11" is to be embedded, $VLC_1$ is replaced with $VLC_4$. To preserve the visual quality, the RSVs corresponding to the VLCs in this mapping set are all modified to the RSV corresponding to $VLC_1$.
\begin{figure}[htbp]
\centering
\includegraphics[width= 3.3 in]{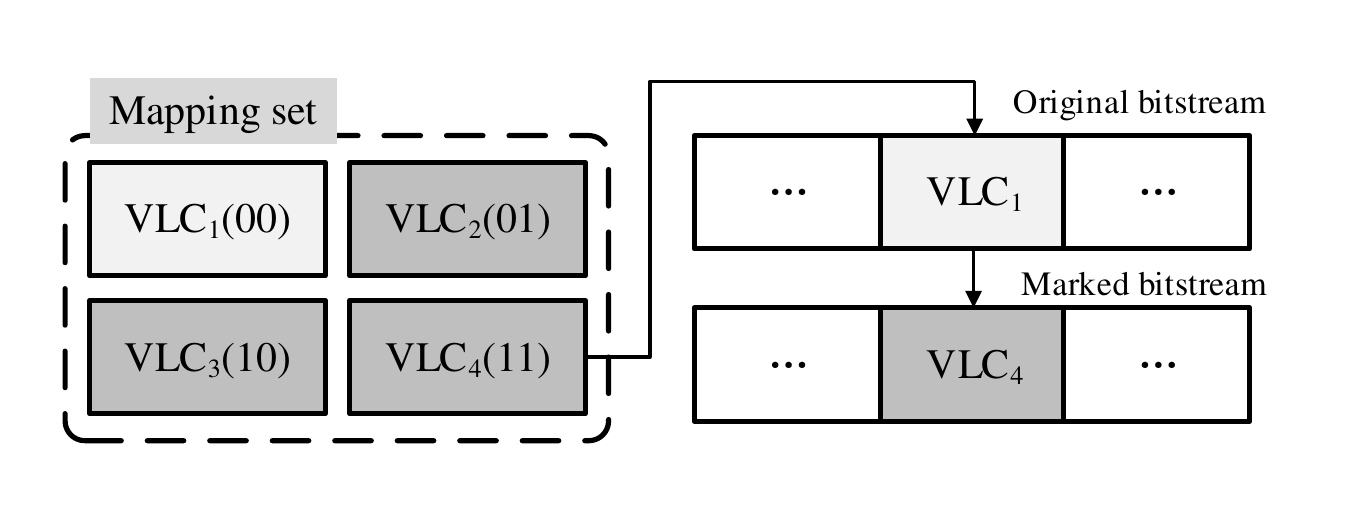}
\caption{The instance of embedding data "11" by replacing the VLC.}
\label{fig_replace}
\end{figure}

\section{Proposed Scheme}
\par In this section, the proposed GVM-based LDH scheme for JPEG bitstream is described in detail. The framework of the proposed scheme is illustrated in Fig. 4. In the preprocessing phase, the VLCs are first parsed and the corresponding RSVs are reordered to generate a sharper histogram of RSVs, which is detailed in subsection \uppercase\expandafter{\romannumeral3}-A. Then the selection of two embedding manners are described in subsection \uppercase\expandafter{\romannumeral3}-B. After determining the embedding manner, for a given payload, the construction of the GVM relationship is detailed in subsection \uppercase\expandafter{\romannumeral3}-C. Secret data is then embedded in the JPEG bitstream using the GVM relationship. Furthermore, for the visual quality of the marked JPEG image with no distortion, the file header is also modified. The detailed embedding process and extraction process are presented in subsection \uppercase\expandafter{\romannumeral3}-D.
\begin{figure*}[htbp]
\centering
\includegraphics[width = 7 in]{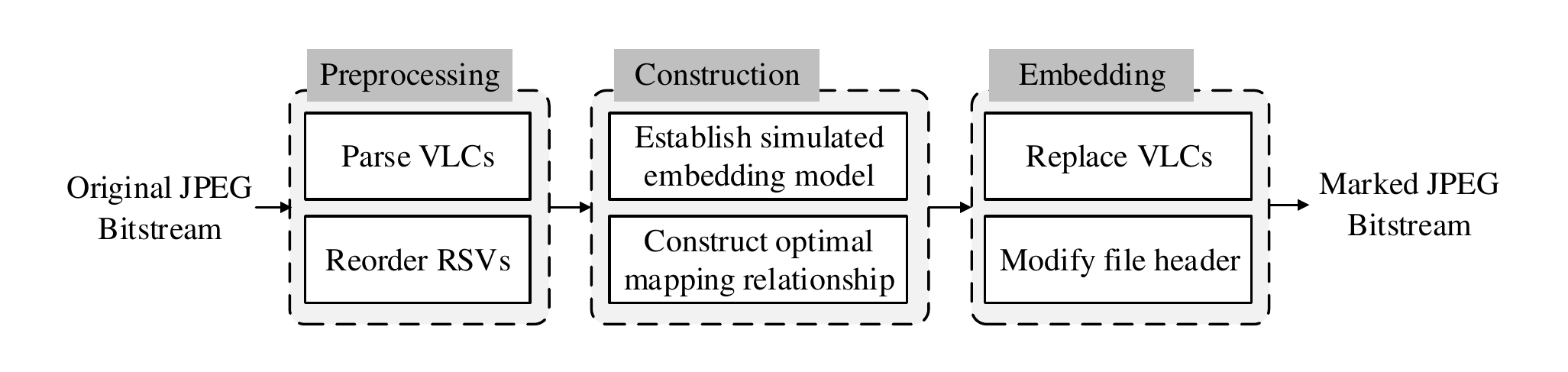}
\caption{The framework of the proposed scheme.}
\label{fig_framework}
\end{figure*}
\subsection{Preprocessing}
\subsubsection{VLCs Parsing}
\par After reading the entropy-coded data, all the frequencies of VLCs are counted. Then the Huffman table are generated by reading the DHT segment. Each RSV in the DHT segment is corresponding to one unique VLC, so the frequencies of VLCs are equivalent to the frequencies of RSVs. We denote the frequency of RSVs and VLCs as ${F}$, such that ${F=\left \{f_{1},f_{2},\cdots,f_{N}\right \}}$, where $f_i$ is the frequency of the ${i}$-th RSV in the DHT segment and $N$ is the number of RSVs. 
\subsubsection{RSVs Reordering}

\par The purpose of reordering the RSVs consists of two points. First, for most JPEG images, the default Huffman table is used to encode the DCT coefficients. However, the generation of the default Huffman table is not according to how frequently each VLC appears in a specific JPEG image. Thus, there may exist coding redundancy when using the default Huffman table. After reordering the RSVs, the coding redundancy can be further reduced. Second, the reduction of coding redundancy is an initialized-liked process. After minimizing the coding redundancy, the performance of arbitrary mapping relationship can be easily measured to obtain the optimal mapping relationships. 
\par According to the Huffman coding, the short VLCs are assigned to the RSVs with high frequency. Therefore, the RSVs are reordered according to the frequencies in the descending order. The frequencies of the reordered RSVs are represented as ${F^{\prime}=\left \{f_{1}^{\prime},f_{2}^{\prime},\cdots,f_{N}^{\prime}\right \}}$. Then the coding redundancy $C$ can be calculated by 
\begin{equation}
\label{eq1}
\centering
\begin{aligned}
C=\sum_{i=1}^{N}\left ( \left ( f_{i}-f_{i}^{\prime} \right )\cdot l_{i} \right ),
\end{aligned}
\end{equation}
where $l_i$ denotes the length of the i-th VLC, in the order of original RSVs. Fig. 5 shows the histograms of the first 30 RSVs before and after reordering for the image Baboon with quality factor (QF) of 70. $X$-axis stands for the index of RSVs in the DHT segment. It can be easily seen from Fig. 5 that the reordered RSV histogram is sharper and suitable for modification.

\subsection{Selection of Embedding Manner}
\par Before presenting our construction of the GVM relationship, we first describe two kinds of embedding manners. Then, which manner is more suitable to construct the GVM relationship is analyzed.
\subsubsection{Direct Mapping (DM) Embedding Manner} 
\par In the DM embedding manner, several unused VLCs, and one used VLC are selected to construct the mapping relationship directly. The positions in the file header of the RSVs corresponding to these unused VLCs are unchanged. An example of the DM embedding manner is illustrated in Fig. 5. Assume that the frequencies of the RSVs after $RSV_4$ are all zero, which is the VLCs corresponding to $RSV_5$ and $RSV_6$ are unused VLCs. Then they are mapped to $RSV_1$ and $RSV_2$ respectively. The other RSVs are not involved in the construction of the mapping relationship. 
\begin{figure}[htbp]
\centering
\includegraphics[width= 3.4 in]{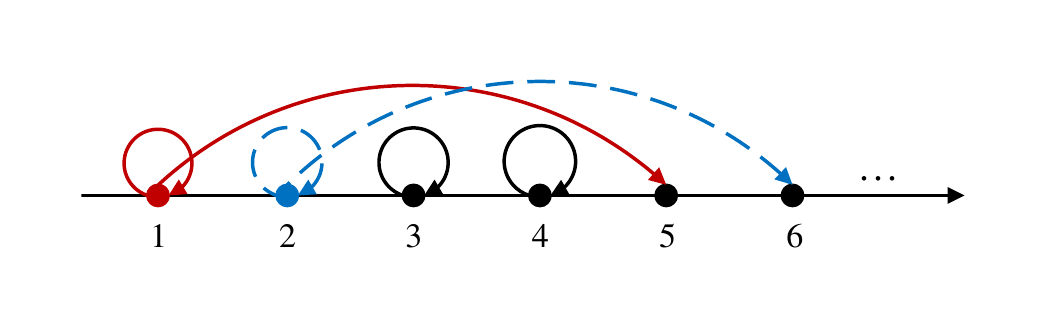}
\caption{Example of the direct mapping embedding manner.}
\label{fig_direct}
\end{figure}

\subsubsection{Histogram Shifting (HS) Embedding Manner} 
\par Based on the histogram of reordered RSVs, the VLC mapping model can be converted to the HS model. For the histogram of RSVs, one RSV corresponding to the used VLC can be seen as one peak bin and one RSV corresponded to the unused VLC can be seen as one zero bin. In the embedding process using this manner, all the bins to the right of the selected peak bin are shifted one or more units. The several nearest bins to the right of the selected peak bin are changed to zero bins. Then the secret data can be embedded by replacing the original VLCs with one of the VLCs in the same mapping set. After the shifting operation, the lengths of VLCs in a mapping set are close or even equal, which can reduce the file size expansion. Multiple peak bins can be selected to obtain higher embedding capacity and better performance of the file size preservation. For one selected peak bin, the number of assigned zero bins is also multiple. Fig. 8 illustrates the HS embedding manner. In Fig. 8, $RSV_1$ is expanded to embed data, $RSV_2$ is firstly shifted to the right by 1 unit and then expanded to embed data. The other RSVs are shifted to the right by 2 units to create vacant bins.

\begin{figure}[htbp]
\centering
\includegraphics[width= 3.4 in]{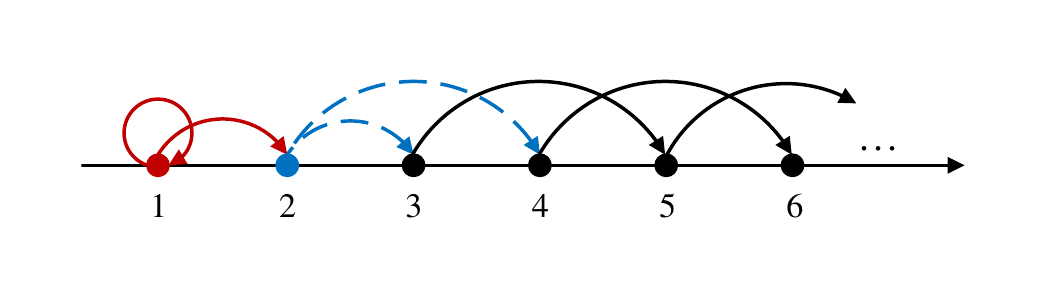}
\caption{Example of histogram shifting embedding manner.}
\label{fig_shifting}
\end{figure}

\subsubsection{Analysis of Two Embedding Manners}
\par In all the previous VLC-based schemes, the DM embedding manner is utilized. Since the used VLC and unused VLCs in the same mapping set are with the same length, the length of the VLC in the JPEG bitstream is unchanged before and after replacement. However, if we use this manner to construct the GVM relationship, the file size of marked JPEG bitstream will be increased significantly than the original JPEG bitstream. After reordering the RSVs in descending order, these zero-frequency RSVs are at the end of the RSV sequence. In general, the lengths of unused VLCs are close or equal to 16 bits. While the lengths of used VLCs corresponding to the high-frequency RSVs are far shorter than 16 bits. The increased file size is unacceptable after replacement. Therefore, we select the HS embedding manner to construct the GVM relationship.
\subsection{Construction}
\par For the secret data with a given payload, much feasible GVM relationships can be constructed according to different parameters. To find the optimal combination of parameters, a simulated embedding model is established to measure the file size increments of each feasible GVM relationship.
\subsubsection{Parameters Definition}
\par In the HS embedding manner, the key issue is the determination of multiple parameters. The parameters consist of the selected peak bins and the numbers of zero bins assigned for all the selected peak bins. Different parameter combinations produce different mapping relationships. One kind of parameter combination is called one feasible solution. In the proposed scheme, some insights on parameter selection are presented to generate fewer parameter combinations. Then we establish a simulated embedding model to measure the simulated file size increments for each feasible solution to find the optimal solution, which are the optimal parameters for the HS manner.
\par First, the histogram of the reordered RSVs $h$ can be represented as:
\begin{equation}
	h(i)=\{f_i^{\prime} \mid 1\leqslant i \leqslant N\},
\end{equation}
where $h(i)$ is the frequency of the $i$-th reordered RSV, which is equal to $f_i^{\prime}$. For the given payload $P$, the extreme feasible solution is that only selecting one peak bin and assigning one zero bin for it while embedding. In this extreme case, to reduce the file size increments while embedding all the secret bits, the frequency of the selected peak bin should be greater than or equal to $P$ and the closest to $P$. Denote the extreme peak bin by $p^{e}$, then the selection of $p^{e}$ is equivalent to solve the following optimization problem:
\begin{equation}
	p^{e}=\mathop{\arg\min}\limits_{i \in N} \{h(i) \mid h(i)\geqslant P \}.
\end{equation}
In our proposed scheme, we only select the peak bins which are greater than or equal to $p^{e}$ to generate a feasible solution. Moreover, all the frequencies of the selected bins should be nonzero, which means that the peak bins should be smaller than or equal to the total number of nonzero bins $N_{nz}$. Thus, the ranges of the starting selected peak bin $p^{s}$, which is on the far left of the selected peak bins, is represented as:
\begin{equation}
	p^{s}=\{i\mid p^{e}\leqslant i\leqslant N_{nz}\},
\end{equation} 
where $N_{nz}$ is the total number of nonzero bins. After determining the starting peak bin $p^s$, the optional number of selected peak bins $m$ is determined by both of the user's preferences and $N_{nz}$:
\begin{equation}
	m=\min\{(N_{nz}-p^{s}+1),U\},
\end{equation}
where $U$ is the optional number of selected peak bins according to user's preferences. Then all the selected peak bins in $h$ can be expressed as:
\begin{equation}
p_n=\{p^s+n-1\mid 1\leqslant n \leqslant m\},
\end{equation}
where $p_n$ is the $n$-th selected peak bin.
\par Different solution spaces can be generated according to different combinations in $p^s$ and $m$. Several constraint conditions are designed to guide the determination of one feasible solution. For one feasible solution, the numbers of zero bins assigned to all the selected peak bins,  $\{a_{1},a_{2},\cdots,a_{m}\}$, should satisfy the following conditions:
\begin{equation}
\small
\left\{
\begin{array}{lll}
\sum_{k=1}^{m}a_{k}\leqslant N_z  & \\
\sum_{k=1}^{m}\log_2(a_{k}+1)\times h(p^k)\geqslant P  & \\
a_{k+1}\leqslant a_{k},& 1\leqslant k< m \\
a_{k}=2^j-1, & j=1,2,\cdots,6
\end{array}
\right.,
\end{equation}
where $N_z$ is the total number of zero bins in the RSV histogram. The first condition is to limit the total number of the assigned zero bins that are less than or equal to the number of zero bins $N_z$. The second condition is to restrict the capacity obtained by this solution should be greater than or equal to the given payload $P$. For the third condition, the number of zero bins assigned to the currently selected peak bin should be less than or equal to that assigned to the previous peak bin, the purpose is to filter the feasible solutions more effectively because the RSVs are reordered in descending order according to the frequency. Furthermore, the number of zero bins should be one of the $\{0,1,3,7,15,31,63\}$ to embed the binary secret bits conveniently. The number of assigned zero bins is zero when the current selected peak bins and after peak bins are not used to carry data.
\subsubsection{Establishment of Simulated Embedding Model}
\par Different feasible solutions can be generated according to different combinations of $(p^s,m,\{a_1,a_2,\cdots,a_m\})$. To find the optimal solution, the generation of complete solution space is required. Then a simulated embedding model is established to measured the performance of each feasible solution. 
\par As mentioned before, the visual quality of the marked JPEG images is no degradation using the proposed scheme. Therefore, we only need to compare the file size increments at the given payload $P$ for each feasible solution. During the shifting operation, the RSVs between the selected peak bin and zero bins are shifting to the right one or more units. The lengths of corresponding VLCs of these RSVs may be increased after shifting, which means that the shifting operation will cause file size expansion. In the embedding process, the original VLC are replaced with the VLC in one mapping set. For the HS embedding manner, the zero bins assigned to the selected bin are to the right of the selected peak bin, which means the lengths of VLCs in one mapping set are greater than or equal to the original VLC, which causes file size expansion. Therefore, we established a simulated embedding model to simulate the shifting and embedding process and the file size increments caused by shifting and embedding can be calculated. An example of simulated embedding is illustrated in Fig. 8. As Fig.8 described, the first selected bin is only expanded to carry data. The second bin is first shifted by 1 unit and then expanded to carry data while the other bins are only shifted by two units to create vacated bins.
\begin{figure*}[htbp]
\centering
\includegraphics[width= 7 in]{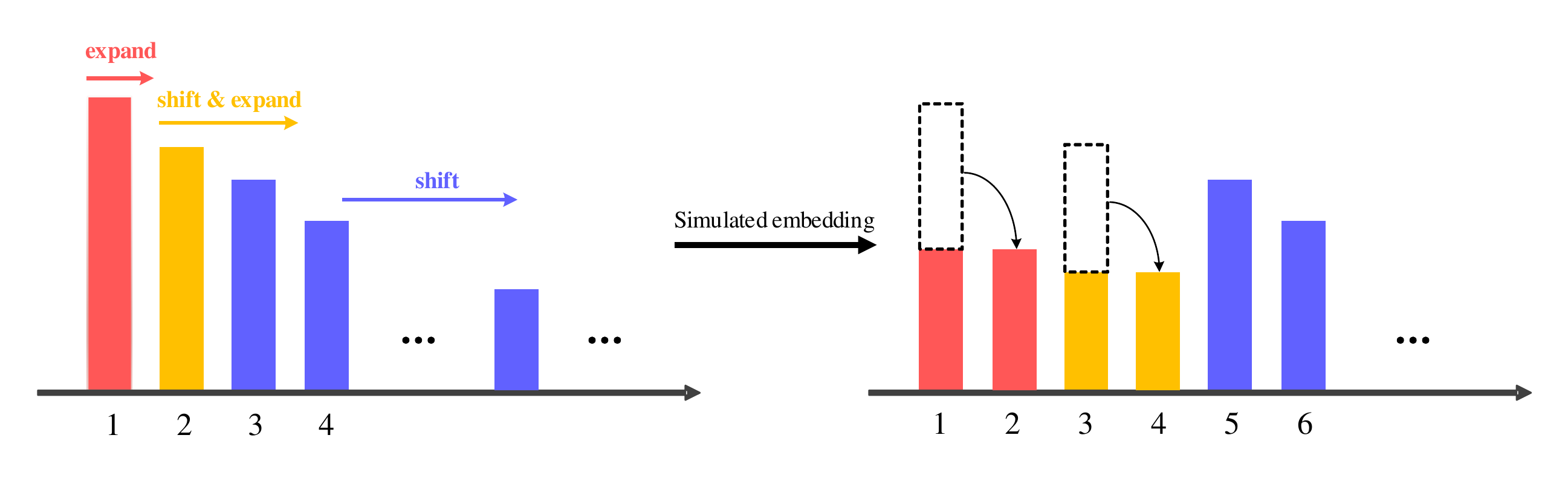}
\caption{Example of simulated embedding with two pairs of peak and zero bins.}
\label{simulated_embedding}
\end{figure*}
\par Assume that one feasible solution is the $j$-th solution in the complete solution space, which is denoted by $(p_j^s,m_j,\{a_{j,1},a_{j,2},\cdots,a_{j,m_j}\})$. All the selected peak bins can be expressed as:
\begin{equation}
p_{j,n}^{\prime}=\left\{
\begin{array}{lll}
p_{j,n},  & \mathrm{if}\ n=1 \\
p_{j,n}+\sum_{k=1}^{n-1}a_{j,k}, & \mathrm{if}\ 2\leqslant n \leqslant m_j\\
\end{array}
\right.,
\end{equation}
where $p_{j,n}$ is the $n$-th selected peak bin. In the shifting process, except for the first selected peak bin $p^s$, other selected peak bins are all shifted to the right by multiple units. For each combination of $\{a_{j,1},a_{j,2},\cdots,a_{j,m_j}\}$, the selected peak bins after shifting can be expressed as:
\begin{equation}
p_{j,n}^{\prime}=\{p_{j,n}-1+\sum_{k=1}^{n-1}a_{j,k}\mid 1\leqslant n \leqslant m_j\},
\end{equation}
where $p_{j,n}^{\prime}$ is the $n$-th selected peak bin after shifting. Then the RSV histogram after simulated shifting, $h_{j}^{s}$, can be represented as:
\begin{eqnarray}
\small
h_{j}^{s}(i)=\left\{
\begin{array}{lll}
h(i),  & \mathrm{if}\ 1\leqslant i< p_s \\
h(p_{j,n}), & \mathrm{if}\ i=p_{j,n}^{\prime}\\
h(i-\sum_{k=1}^{m}a_{j,k}),  & \mathrm{if}\ p_{j,m}^{\prime}+a_{j,m}+1 \leqslant i \\
&  \leqslant N_{nz}+\sum_{k=1}^{m}a_{j,k}                        \\
0, & \mathrm{others}
\end{array}
\right.,
\end{eqnarray}
where $h_{j}^{s}(i)$ is the frequency of $i$-th bin of $h_{j}^{s}$. Then the simulated file size increments $SI_j^s$ caused by shifting can be calculated directly as
\begin{eqnarray}
SI_j^s=\sum_{i=1}^{N}(h^{s}_j(i)-h(i))\times l_i,
\end{eqnarray}
where $l_i$ is the length of VLC corresponding the $i$-th RSV.
\par Generally, the secret data before embedding should be encrypted to improve security. The secret data after encrypting satisfies $independently\ and\ identically\ distributed$ (i.i.d.), so the frequency of each VLC in a mapping set used to embed can be seen as equal. Thus, the RSV histogram after simulated embedding, $h_{j}^{e}$, can be represented as:
\begin{eqnarray}
\small
h^{e}_j(i)=\left\{
\begin{array}{lll}
round(\frac{h^{s}_j(p_{j,n}^{\prime})}{a_{j,n}+1}),  & \mathrm{if}\ p_{j,n}^{\prime}\leqslant i\leqslant p_{j,n}^{\prime}+a_{j,n} \\
h^{s}_j(i), & \mathrm{others}
\end{array}
\right.,
\end{eqnarray}
where $round(\cdot)$ is the rounding function. Then the simulated file size increments $SI_j^e$ caused by embedding can be calculated as
\begin{eqnarray}
SI_j^e=\sum_{i=1}^{N}(h^{e}_j(i)-h^{s}_j(i))\times l_i.
\end{eqnarray}
\par Finally, the total simulated file size increments $SI_j$ can be calculated as
\small
\begin{eqnarray}
\begin{aligned}
SI_j &  = SI_j^s+SI_j^e \\
& =\sum_{i=1}^{N}(h^{s}_i(i)-h(i))\times l_i+ \sum_{i=1}^{N}(h^{e}_j(i)-h^{s}_j(i))\times l_i\\
& =\sum_{i=1}^{N}(h^{e}_i(i)-h(i))\times l_i.
\end{aligned}
\end{eqnarray}
\subsubsection{Construction of GVM Relationship}
\par The feasible solution with the smallest simulated file size increments is the optimal solution which used to construct the GVM relationship. The constructed optimal GVM relationship $G_{opt}$ can be described as Eq.(\ref{GVM}). In Eq.(\ref{GVM}), each mapping set consists of the VLCs corresponding to one selected peak bin and several assigned zero bins.
\begin{equation}
\footnotesize
    G_{opt}=
    \begin{Bmatrix}
        \begin{aligned}
            & \left \{ VLC_{p_1^{\prime}}\leftrightarrow \left \{ VLC_{p_1^{\prime}+1},\cdots ,VLC_{p_1^{\prime}+a_1} \right \} \right \},\\ 
            & \left \{ VLC_{p_2^{\prime}}\leftrightarrow \left \{ VLC_{p_2^{\prime}+1},\cdots ,VLC_{p_2^{\prime}+a_2} \right \} \right \},\cdots,\\ 
            & \left \{ VLC_{p_m^{\prime}}\leftrightarrow \left \{ VLC_{p_m^{\prime}+1},\cdots ,VLC_{p_m^{\prime}+a_m} \right \} \right \}  
        \end{aligned}
    \end{Bmatrix}.
    \label{GVM}
\end{equation}

\subsection{Data Embedding \& Extraction}
\par After the optimal GVM relationship $G_{opt}$ constructed, the secret data can be embedded by replacing the original VLC with the mapped VLC. To keep the marked JPEG image quality has good format compatibility and no degradation, we modify the RSVs sequence in the DHT segment according to $G_{opt}$. As mentioned in section \uppercase\expandafter{\romannumeral2}, each RSV corresponds to one unique VLC. While in a mapping set, the frequency of the unused VLCs are zero, so the corresponding RSVs can be changed to the RSV corresponding to the used VLC. Then the RSVs corresponding to all the VLCs in one mapping set is the same, which means that the RSV is unchanged before and after embedding. Therefore, the visual quality is preserved and no error will occur during the decoding process.
\par When embedding the secret data, each VLC is read in the entropy-coded data successively. For the current VLC, we firstly judge whether the corresponding RSV appears multiple times in the modified DHT segment or not. Appear multiple times means that the VLC belongs to one mapping set. If yes, we replace the VLC with another VLC according to the data to be embedded. In general, the Huffman table adopted by the original JPEG image is the default Huffman table, which can be obtained on the Internet. Thus, there are no auxiliary information to be embedded.
\par Based on the above discussion, the detailed data embedding steps are summarized as:

\begin{enumerate}[Step 1.]

\item Decode the original JPEG bitstream $J$ and parse all the VLCs according to the Huffman table in the file header. Then reorder the corresponding RSVs in descending order.

\item Generate the complete feasible solution space according to \uppercase\expandafter{\romannumeral3}-C.

\item According to the simulated embedding model, calculate the simulated file size increments for each feasible solution and then find the optimal solution.

\item Construct the optimal GVM relationship $G_{opt}$ by modifying the corresponding RSVs in the file header. 

\item Replacing the original VLCs with the VLCs in the same mapping set according to the current data to be embedded.

\item Repeat Step 5 until all the data is embedded and the marked JPEG bitstream $J_M$ is generated.

\end{enumerate}

\par The detailed data extraction steps are summarized as follows:

\begin{enumerate}[Step 1.]

\item Decode the marked JPEG image $J_M$ and parse all the VLCs from the entropy-coded data.

\item Restore the adopted GVM relationship $G_{opt}$ according to the modified DHT segment in the file header.

\item Extract the embedded data and restore the entropy-coded data according to the GVM relationship $G_{opt}$.     

\item Restore the original RSVs in the file header, then the original JPEG bitstream $J$ is restored. 

\end{enumerate}

\section{Experimental results}
\begin{figure*}[htbp]
\centering
    \subfigure[Baboon]{
    \includegraphics[width= 1.1 in]{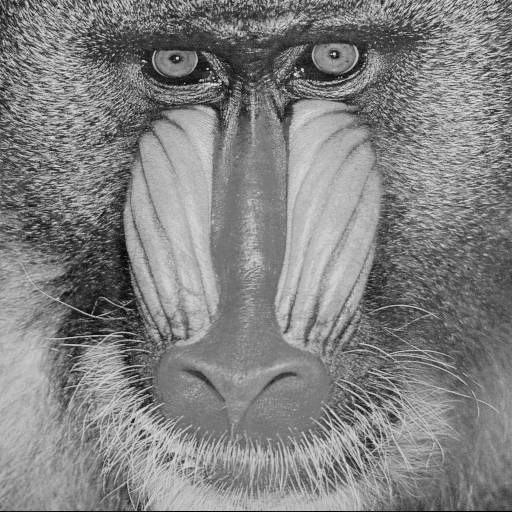}}
    \subfigure[Boat]{
    \includegraphics[width= 1.1 in]{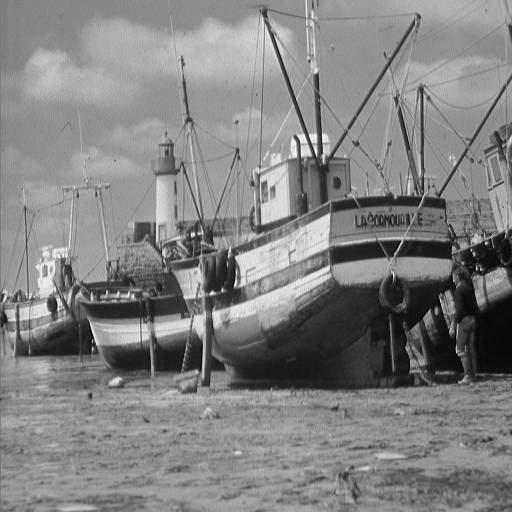}}
    \subfigure[Elaine]{
    \includegraphics[width= 1.1 in]{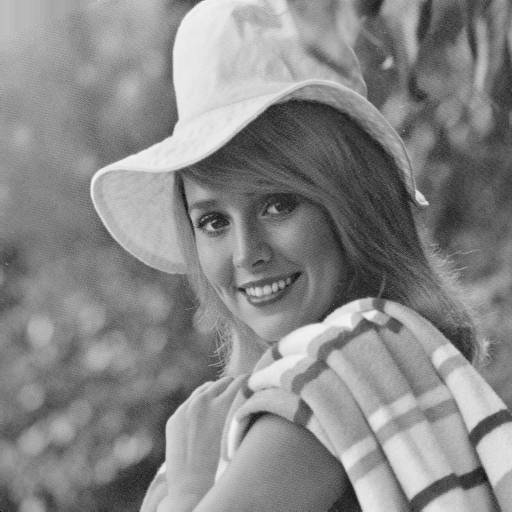}}
    \subfigure[Lena]{
    \includegraphics[width= 1.1 in]{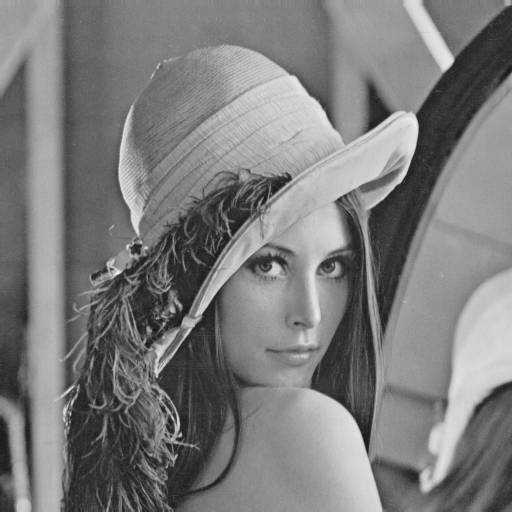}}
    \subfigure[Peppers]{
    \includegraphics[width= 1.1 in]{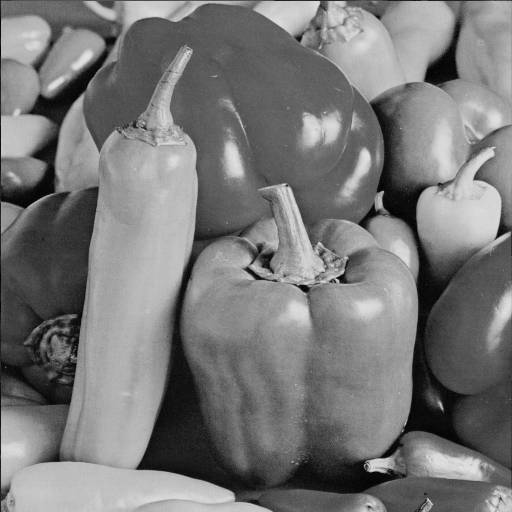}}
    \subfigure[Tiffany]{
    \includegraphics[width= 1.1 in]{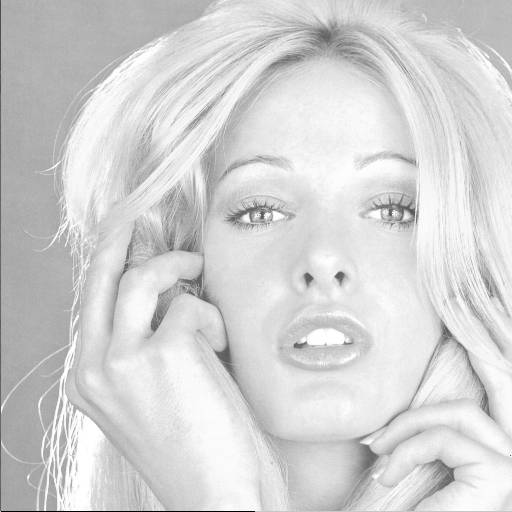}}
\caption{The test images with QF = 70.}
\label{testimages}
\end{figure*}

\par In our experiments, six 512$\times$512$\times$8 bits grayscale test images, including Baboon, Lena, Tiffany, Peppers, Elaine and Airplane, from the USC-SIPI database \cite{IEEEexample:usc-sipi} are converted into grayscale JPEG image with various QFs to test. Fig. 9 illustrates these test images after converted with QF =70. Moreover, to investigate the suitability of our proposed scheme, 200 images are randomly selected from the image database UCID \cite{IEEEexample:schaefer2003ucid}. The selected 200 images are converted into grayscale JPEG images with four QFs including 30,50,70, and 90. In the proposed scheme, the default Huffman table is modified to construct GVM relationship. Therefore, all the test JPEG images are compressed with the default Huffman table by using the IJG toolbox \cite{IJG_toolbox}.
\par To evaluate the performance of our proposed scheme, serval previous VLC-based schemes are selected for comparison, which are proposed by including Qian and Zhang \cite{IEEEexample:qian2012lossless}, Hu $ et~al. $\cite{IEEEexample:hu2013improved}, and Qiu $ et~al. $\cite{IEEEexample:qiu2018lossless}. Several state-of-the-art DCT-based RDH schemes are also selected for comparison, which are proposed by including Huang ${ et~al. }$\cite{IEEEexample:huang2016reversible}, Qian ${ et~al. }$\cite{IEEEexample:qian2017reversible}, Wedaj ${ et~al. }$\cite{IEEEexample:wedaj2017improved}, and Hou ${ et~al. }$\cite{IEEEexample:hou2018reversible} respectively. Several experiments are conducted to demonstrate the effectiveness of the proposed scheme. Given a test JPEG image, we investigate two aspects, which are visual quality and file size preservation. The performance of visual quality is measured using the peak signal-to-noise ratio (PSNR) value between the cover JPEG image and the marked JPEG image. The performance of file size preservation is measured by the file size increments between the marked JPEG bitstream and the cover JPEG bitstream.

\subsection{Comparison with VLC-based schemes}

\begin{table*}[htbp]
\centering
\caption{Embedding capacity and file size increments comparison with previous VLC-based schemes for different test images at different QFs}
\label{tab:capacity}
\scriptsize
\begin{tabular}{@{}ccrrrrrrrr@{}}
\toprule
\multicolumn{1}{c}{\multirow{2}{*}{Image}} & \multirow{2}{*}{Scheme} & \multicolumn{2}{c}{QF=30}                                         & \multicolumn{2}{c}{QF=50}                                         & \multicolumn{2}{c}{QF=70}                                         & \multicolumn{2}{c}{QF=90}                                         \\ \cmidrule(l){3-10} 
\multicolumn{1}{c}{}                       &                         & \multicolumn{1}{c}{Capacity(bits)} & \multicolumn{1}{c}{FI(bits)} & \multicolumn{1}{c}{Capacity(bits)} & \multicolumn{1}{c}{FI(bits)} & \multicolumn{1}{c}{Capacity(bits)} & \multicolumn{1}{c}{FI(bits)} & \multicolumn{1}{c}{Capacity(bits)} & \multicolumn{1}{c}{FI(bits)} \\ \midrule
Baboon                                     & Qian and Zhang\cite{IEEEexample:qian2012lossless}          & 1592                               & 0                            & 1117                               & 0                            & 1089                               & 0                            & 342                                & 0                            \\
                                          & Hu $et~al.$\cite{IEEEexample:hu2013improved}           & 1739                               & 0                            & 1254                               & 0                            & 1348                               & 0                            & 662                                & 0                            \\
                                          & Qiu $ et~al. $\cite{IEEEexample:qiu2018lossless}          & 1966                               & 0                            & 1473                               & 0                            & 1620                               & 0                            & 926                                & 0                            \\
                                          & Proprosed(1,1,\{1\})    & 16335                              & 12093                        & 20588                              & 23586                        & 25251                              & 36616                        & 38776                              & 57807                        \\
                                          & Proposed(1,2,\{1,1\})   & 26279                              & 23712                        & 34687                              & 39062                        & 43548                              & 57727                        & 67540                              & 94193                        \\ \midrule
Boat                                       & Qian and Zhang \cite{IEEEexample:qian2012lossless}         & 649                                & 0                            & 552                                & 0                            & 428                                & 0                            & 978                                & 0                            \\
                                          & Hu $ et~al. $ \cite{IEEEexample:hu2013improved}          & 685                                & 0                            & 681                                & 0                            & 667                                & 0                            & 1935                               & 0                            \\
                                          & Qiu $ et~al. $ \cite{IEEEexample:qiu2018lossless}         & 1242                               & 0                            & 1079                               & 0                            & 951                                & 0                            & 2165                               & 0                            \\
                                          & Proprosed(1,1,\{1\})    & 8966                               & 6936                         & 11686                              & 12829                        & 14809                              & 20499                        & 25553                              & 35019                        \\
                                          & Proposed(1,2,\{1,1\})   & 13671                              & 12887                        & 18835                              & 22244                        & 24758                              & 33161                        & 42438                              & 59292                        \\ \midrule
Elaine                                     & Qian and Zhang\cite{IEEEexample:qian2012lossless}          & 197                                & 0                            & 483                                & 0                            & 326                                & 0                            & 1002                               & 0                            \\
                                          & Hu $ et~al. $\cite{IEEEexample:hu2013improved}           & 198                                & 0                            & 547                                & 0                            & 401                                & 0                            & 1718                               & 0                            \\
                                          & Qiu $ et~al. $\cite{IEEEexample:qiu2018lossless}          & 842                                & 0                            & 974                                & 0                            & 781                                & 0                            & 1900                               & 0                            \\
                                          & Proprosed(1,1,\{1\})    & 5910                               & 3046                         & 8730                               & 7225                         & 13978                              & 12327                        & 31273                              & 33596                        \\
                                          & Proposed(1,2,\{1,1\})   & 10006                              & 7278                         & 13345                              & 13653                        & 20740                              & 22771                        & 49272                              & 56577                        \\ \midrule
Lena                                       & Qian and Zhang\cite{IEEEexample:qian2012lossless}          & 253                                & 0                            & 365                                & 0                            & 214                                & 0                            & 298                                & 0                            \\
                                          & Hu $ et~al. $\cite{IEEEexample:hu2013improved}           & 260                                & 0                            & 367                                & 0                            & 280                                & 0                            & 576                                & 0                            \\
                                          & Qiu $ et~al. $\cite{IEEEexample:qiu2018lossless}          & 916                                & 0                            & 996                                & 0                            & 759                                & 0                            & 841                                & 0                            \\
                                          & Proprosed(1,1,\{1\})    & 6087                               & 4039                         & 8284                               & 8780                         & 11370                              & 14134                        & 21938                              & 29911                        \\
                                          & Proposed(1,2,\{1,1\})   & 10183                              & 8820                         & 12923                              & 15303                        & 17952                              & 23868                        & 35282                              & 48214                        \\ \midrule
Peppers                                    & Qian and Zhang\cite{IEEEexample:qian2012lossless}          & 622                                & 0                            & 433                                & 0                            & 323                                & 0                            & 762                                & 0                            \\
                                          & Hu $ et~al. $\cite{IEEEexample:hu2013improved}           & 707                                & 0                            & 548                                & 0                            & 449                                & 0                            & 1576                               & 0                            \\
                                          & Qiu $ et~al. $\cite{IEEEexample:qiu2018lossless}          & 1077                               & 0                            & 924                                & 0                            & 768                                & 0                            & 1675                               & 0                            \\
                                          & Proprosed(1,1,\{1\})    & 5893                               & 4125                         & 8409                               & 8900                         & 12238                              & 14039                        & 24068                              & 26481                        \\
                                          & Proposed(1,2,\{1,1\})   & 9989                               & 8903                         & 12981                              & 15456                        & 18711                              & 23906                        & 38818                              & 46290                        \\ \midrule
Tiffany                                    & Qian and Zhang\cite{IEEEexample:qian2012lossless}          & 613                                & 0                            & 573                                & 0                            & 816                                & 0                            & 258                                & 0                            \\
                                          & Hu $ et~al. $\cite{IEEEexample:hu2013improved}           & 674                                & 0                            & 775                                & 0                            & 1003                               & 0                            & 765                                & 0                            \\
                                          & Qiu $ et~al. $\cite{IEEEexample:qiu2018lossless}          & 1125                               & 0                            & 1090                               & 0                            & 1284                               & 0                            & 1004                               & 0                            \\
                                          & Proprosed(1,1,\{1\})    & 5499                               & 1403                         & 7815                               & 6149                         & 11580                              & 11252                        & 22906                              & 30523                        \\
                                          & Proposed(1,2,\{1,1\})   & 9595                               & 5305                         & 12098                              & 11857                        & 17823                              & 20304                        & 37065                              & 48196                        \\ \bottomrule
\end{tabular}%
\end{table*}

\par For the previous VLC-based schemes, the lengths of used VLC and the mapped unused VLCs should be the same. The VLCs of different lengths can be classified to 16 categories. While for the most categories, the number of unused VLCs is quite low and even zero. Fig. 10 describes the distribution of VLCs for $Baboon$ with different QFs. The blue histogram shows the frequency of VLCs in different lengths. The orange line chart is the number of unused VLCs in different lengths. As Fig. 10 illustrated, for the categories with a high frequency which are also the VLCs of short length, the number of unused VLCs is quite low and close to zero. While for the categories with low frequency, which is the VLCs with a length of 16, the number of unused VLCs is much high. This is consistent with the characteristics of Huffman coding. Most VLCs with short lengths are used so the number of unused VLCs of short length is close or equal to zero. Therefore, for the previous VLC-based schemes, the embedding capacity is difficult to be improved significantly.
\par As mentioned in Section \uppercase\expandafter{\romannumeral4}-A, for a JPEG image compressed by using the default Huffman table, the coding redundancy exists since the original RSVs are not in a descending order strictly. The coding redundancy for the six test images with various QFs are described in . For one test image, the coding redundancy at different QFs are different and high, which demonstrates the default Huffman table is not the most suitable no matter which the QFs. Furthermore, the lowest redundancy is often obtained with the higher QFs, such as 70, 80, and 90, which indicates that at the high QFs, the frequency order of the original RSVs in JPEG bitstream is more close to the order defined by the default Huffman table. 
\par For a fair comparison, the file size increments caused by shifting and embedding are listed, which called grass file size increments(GFI). Note that, the actual file size increments of our proposed scheme are smaller than GFI because the coding redundancy is subtracted before embedding, which called pure file size increments(PFI). In our proposed scheme, the embedding capacity is different using the different GVM relationships. The GVM relationship is determined by multiple parameters, such as the starting peak bin $p_s$, the number of selected peak bins $m$, and the numbers of zero bins assigned to each selected bin $\{a_1,a_2,\cdots,a_m\}$. To compare the embedding capacity with the previous VLC-based schemes, we designed two combinations of the three parameters. The first combination of the three parameters is $(1,1,\{1\})$. That is only one peak bin is selected, which is on the far left, and the number of zero bins assigned to it is 1. The second combination of the three parameters are $(1,2,\{1,1\})$. That is two peak bins are selected, which are on the far left, and the numbers of zero bins assigned to it are all 1. Table \uppercase\expandafter{\romannumeral1} lists the comparison of embedding capacity and file size increments with previous VLC-based schemes. As Table \uppercase\expandafter{\romannumeral1} illustrates, the embedding capacity of previous VLC-based schemes is rather low. By slightly increasing the file size, the embedding capacity is improved significantly. It can be easily seen from Table \uppercase\expandafter{\romannumeral1} that the PFI is smaller than the capacity in some cases, which is because the high coding redundancy.

\subsection{Comparison with DCT-based schemes}
\par The comparison with the DCT-based RHD schemes consists of two aspects, visual quality and file size preservation, which are measured by PSNR values and file size increments respectively. 
\par Limited to the embedding capacity, we set different payloads for the test images with different QFs. In our experiments, 3000, 6000, 9000, and 12000 bits of additional data are embedded into six test JPEG images with QF = 30. 4000, 8000, 12000, and 16000 bits of additional data are embedded into six test images with QF = 50. 5000, 10000, 15000, and 20000 bits of additional data are embedded into six test images with QF = 70. 6000, 12000, 18000, and 24000 bits of additional data are embedded into six test images with QF = 50. We set the max number of selected peak bins of our proposed scheme is 5 to compare. For the six test images, the PSNR comparison at different payloads are listed in Table \uppercase\expandafter{\romannumeral2}. Table \uppercase\expandafter{\romannumeral2} shows that the PSNR values of our scheme are always $+\infty$. That means our proposed scheme can keep the visual quality unchanged, which is a truly lossless data hiding scheme.

\begin{table*}[htbp]
\centering
\caption{The PSNR values (dB) of the marked JPEG images with different QFs and embedded with different lengths of data using our scheme and four previous schemes.'-' means the capacity is not enough for the payload.}
\label{tab:psnr}
\resizebox{\textwidth}{!}{%
\begin{tabular}{@{}ccrrrrrrrrrrrrrrrr@{}}
\toprule
\multirow{2}{*}{\textbf{Image}} & \multirow{2}{*}{\textbf{Scheme}} & \multicolumn{4}{l}{\textbf{Payload(bits) with QF=30}} & \multicolumn{4}{l}{\textbf{Payload(bits) with QF=50}} & \multicolumn{4}{l}{\textbf{Payload(bits) with QF=70}} & \multicolumn{4}{l}{\textbf{Payload(bits) with QF=90}} \\ \cmidrule(l){3-18} 
                                &                                  & 3000        & 6000        & 9000        & 12000       & 4000        & 8000        & 12000       & 16000       & 5000        & 10000       & 15000       & 20000       & 6000        & 12000       & 18000       & 24000       \\ \midrule
Baboon                          & Huang $et~al.$\cite{IEEEexample:huang2016reversible}                            & 41.96       & 37.78       & 35.11       & 33.22       & 43.65       & 39.41       & 36.43       & 34.07       & 45.33       & 40.74       & 37.75       & 35.16       & 48.25       & 44.17       & 41.39       & 39.33       \\
                                & Qian $et~al.$\cite{IEEEexample:qian2017reversible}                             & 41.63       & 37.81       & 35.16       & 33.23       & 43.69       & 39.37       & 36.69       & 34.21       & 45.42       & 40.94       & 37.82       & 35.18       & 48.71       & 44.24       & 41.50       & 39.33       \\
                                & Wedaj $et~al.$\cite{IEEEexample:wedaj2017improved}                            & 40.55       & 36.97       & 34.78       & 33.19       & 41.97       & 38.39       & 36.11       & 34.44       & 42.86       & 39.48       & 37.54       & 35.71       & 46.25       & 43.23       & 40.83       & 39.33       \\
                                & Hou $et~al.$\cite{IEEEexample:hou2018reversible}                              & 42.23       & 38.12       & 35.50       & 33.55       & 43.94       & 39.84       & 36.92       & 34.50       & 45.73       & 41.23       & 37.99       & 35.42       & 48.52       & 44.34       & 41.44       & 39.35       \\
                                & Proposed ($m$=5)                             & \textbf{$+\infty$}          & \textbf{$+\infty$}          & \textbf{$+\infty$}          & \textbf{$+\infty$}          & \textbf{$+\infty$}          & \textbf{$+\infty$}          & \textbf{$+\infty$}          & \textbf{$+\infty$}          & \textbf{$+\infty$}          & \textbf{$+\infty$}          & \textbf{$+\infty$}          & \textbf{$+\infty$}          & \textbf{$+\infty$}          & \textbf{$+\infty$}          & \textbf{$+\infty$}          & \textbf{$+\infty$}          \\ \midrule
Boat                            & Huang $et~al.$\cite{IEEEexample:huang2016reversible}                            & 41.66       & 37.82       & 35.46       & 33.44       & 43.78       & 39.74       & 37.27       & 35.05       & 46.37       & 42.15       & 39.47       & 37.08       & 50.38       & 46.41       & 44.11       & 42.45       \\
                                & Qian $et~al.$\cite{IEEEexample:qian2017reversible}                             & 41.71       & 37.94       & 35.42       & 33.49       & 44.04       & 39.82       & 37.31       & 35.19       & 46.63       & 42.03       & 39.49       & 37.11       & 51.01       & 46.95       & 44.48       & 42.78       \\
                                & Wedaj $et~al.$\cite{IEEEexample:wedaj2017improved}                            & 41.25       & 37.70       & 35.71       & 33.91       & 42.98       & 39.49       & 37.41       & 35.57       & 45.87       & 41.90       & 39.57       & 37.82       & 50.63       & 47.13       & 45.44       & 43.52       \\
                                & Hou $et~al.$\cite{IEEEexample:hou2018reversible}                              & 42.28       & 38.40       & 35.89       & 33.83       & 44.51       & 40.46       & 37.73       & 35.30       & 47.13       & 42.76       & 39.91       & 37.27       & 50.82       & 46.64       & 44.17       & 42.45       \\
                                & Proposed ($m$=5)                             & \textbf{$+\infty$}          & \textbf{$+\infty$}          & \textbf{$+\infty$}          & \textbf{$+\infty$}          & \textbf{$+\infty$}          & \textbf{$+\infty$}          & \textbf{$+\infty$}          & \textbf{$+\infty$}          & \textbf{$+\infty$}          & \textbf{$+\infty$}          & \textbf{$+\infty$}          & \textbf{$+\infty$}          & \textbf{$+\infty$}          & \textbf{$+\infty$}          & \textbf{$+\infty$}          & \textbf{$+\infty$}          \\ \midrule
Elaine                          & Huang $et~al.$\cite{IEEEexample:huang2016reversible}                            & 42.54       & 39.08       & 36.67       & ‒           & 44.55       & 41.15       & 39.11       & 37.03       & 46.56       & 43.22       & 41.13       & 39.44       & 50.15       & 45.92       & 43.36       & 41.53       \\
                                & Qian $et~al.$\cite{IEEEexample:qian2017reversible}                             & 42.55       & 38.89       & 36.64       & ‒           & 44.69       & 41.40       & 39.14       & 37.09       & 46.98       & 43.58       & 41.43       & 39.75       & 51.13       & 46.79       & 43.87       & 42.00       \\
                                & Wedaj $et~al.$\cite{IEEEexample:wedaj2017improved}                            & 42.69       & 38.84       & 36.61       & ‒           & 45.71       & 41.37       & 39.35       & 37.19       & 47.94       & 44.02       & 42.14       & 40.26       & 50.68       & 48.00       & 46.01       & 44.57       \\
                                & Hou $et~al.$\cite{IEEEexample:hou2018reversible}                              & 43.39       & 39.21       & 36.65       & ‒           & 45.77       & 41.83       & 39.26       & 37.11       & 47.72       & 44.08       & 41.64       & 39.72       & 51.07       & 46.50       & 43.65       & 41.84       \\
                                & Proposed ($m$=5)                             & \textbf{$+\infty$}          & \textbf{$+\infty$}          & \textbf{$+\infty$}          & \textbf{$+\infty$}          & \textbf{$+\infty$}          & \textbf{$+\infty$}          & \textbf{$+\infty$}          & \textbf{$+\infty$}          & \textbf{$+\infty$}          & \textbf{$+\infty$}          & \textbf{$+\infty$}          & \textbf{$+\infty$}          & \textbf{$+\infty$}          & \textbf{$+\infty$}          & \textbf{$+\infty$}          & \textbf{$+\infty$}          \\ \midrule
Lena                            & Huang $et~al.$\cite{IEEEexample:huang2016reversible}                            & 42.45       & 38.13       & 35.16       & ‒           & 45.18       & 40.56       & 37.33       & ‒           & 48.02       & 43.83       & 40.51       & 37.26       & 53.48       & 49.84       & 47.50       & 45.20       \\
                                & Qian $et~al.$\cite{IEEEexample:qian2017reversible}                             & 42.47       & 38.14       & 35.00       & ‒           & 45.20       & 40.59       & 37.37       & ‒           & 48.06       & 43.77       & 40.52       & 37.26       & 53.75       & 50.20       & 47.77       & 45.24       \\
                                & Wedaj $et~al.$\cite{IEEEexample:wedaj2017improved}                            & 40.99       & 37.92       & 35.17       & ‒           & 44.40       & 40.27       & 37.49       & ‒           & 46.75       & 42.83       & 40.45       & 37.32       & 52.82       & 48.99       & 47.15       & 45.27       \\
                                & Hou $et~al.$\cite{IEEEexample:hou2018reversible}                              & 42.57       & 38.35       & 35.21       & ‒           & 45.41       & 40.86       & 37.47       & ‒           & 48.32       & 43.97       & 40.70       & 37.26       & 53.99       & 50.17       & 47.56       & 45.24       \\
                                & Proposed ($m$=5)                             & \textbf{$+\infty$}          & \textbf{$+\infty$}          & \textbf{$+\infty$}          & \textbf{$+\infty$}          & \textbf{$+\infty$}          & \textbf{$+\infty$}          & \textbf{$+\infty$}          & \textbf{$+\infty$}          & \textbf{$+\infty$}          & \textbf{$+\infty$}          & \textbf{$+\infty$}          & \textbf{$+\infty$}          & \textbf{$+\infty$}          & \textbf{$+\infty$}          & \textbf{$+\infty$}          & \textbf{$+\infty$}          \\ \midrule
Peppers                         & Huang $et~al.$\cite{IEEEexample:huang2016reversible}                            & 42.56       & 37.72       & 34.92       & ‒           & 45.20       & 40.96       & 37.65       & ‒           & 47.98       & 43.99       & 41.24       & 38.39       & 51.94       & 48.11       & 45.64       & 43.57       \\
                                & Qian $et~al.$\cite{IEEEexample:qian2017reversible}                             & 42.25       & 37.70       & 35.00       & ‒           & 45.07       & 41.08       & 37.68       & ‒           & 47.82       & 43.94       & 41.31       & 38.48       & 52.95       & 49.39       & 46.79       & 44.46       \\
                                & Wedaj $et~al.$\cite{IEEEexample:wedaj2017improved}                            & 41.75       & 37.68       & 35.27       & ‒           & 44.35       & 40.85       & 38.00       & ‒           & 47.63       & 43.87       & 41.05       & 39.01       & 51.80       & 49.23       & 47.20       & 45.47       \\
                                & Hou $et~al.$\cite{IEEEexample:hou2018reversible}                              & 42.64       & 38.49       & 35.26       & ‒           & 45.91       & 41.48       & 38.22       & ‒           & 48.63       & 44.65       & 41.72       & 38.56       & 53.14       & 48.84       & 46.00       & 43.84       \\
                                & Proposed ($m$=5)                             & \textbf{$+\infty$}          & \textbf{$+\infty$}          & \textbf{$+\infty$}          & \textbf{$+\infty$}          & \textbf{$+\infty$}          & \textbf{$+\infty$}          & \textbf{$+\infty$}          & \textbf{$+\infty$}          & \textbf{$+\infty$}          & \textbf{$+\infty$}          & \textbf{$+\infty$}          & \textbf{$+\infty$}          & \textbf{$+\infty$}          & \textbf{$+\infty$}          & \textbf{$+\infty$}          & \textbf{$+\infty$}          \\ \midrule
Tiffany                         & Huang $et~al.$\cite{IEEEexample:huang2016reversible}                            & 42.59       & 37.99       & 35.46       & ‒           & 45.09       & 40.31       & 37.48       & ‒           & 48.24       & 43.29       & 40.38       & 37.62       & 53.71       & 50.09       & 47.39       & 44.91       \\
                                & Qian $et~al.$\cite{IEEEexample:qian2017reversible}                             & 42.65       & 38.00       & 35.56       & ‒           & 45.23       & 40.62       & 37.75       & ‒           & 48.27       & 43.46       & 40.64       & 37.78       & 53.73       & 50.28       & 47.63       & 44.99       \\
                                & Wedaj $et~al.$\cite{IEEEexample:wedaj2017improved}                            & 41.81       & 38.34       & 35.98       & ‒           & 44.82       & 40.69       & 38.25       & ‒           & 47.12       & 43.53       & 41.20       & 38.13       & 51.81       & 49.10       & 47.57       & 45.51       \\
                                & Hou $et~al.$\cite{IEEEexample:hou2018reversible}                              & 42.75       & 38.80       & 35.68       & ‒           & 45.49       & 41.16       & 37.78       & ‒           & 48.73       & 44.18       & 40.84       & 37.69       & 54.30       & 50.61       & 47.70       & 45.01       \\
                                & Proposed ($m$=5)                             & \textbf{$+\infty$}          & \textbf{$+\infty$}          & \textbf{$+\infty$}          & \textbf{$+\infty$}          & \textbf{$+\infty$}          & \textbf{$+\infty$}          & \textbf{$+\infty$}          & \textbf{$+\infty$}          & \textbf{$+\infty$}          & \textbf{$+\infty$}          & \textbf{$+\infty$}          & \textbf{$+\infty$}          & \textbf{$+\infty$}          & \textbf{$+\infty$}          & \textbf{$+\infty$}          & \textbf{$+\infty$}          \\ \bottomrule
\end{tabular}%
}
\end{table*}

\begin{table*}[htbp]
\centering
\caption{The file size increments (bits) of the marked JPEG images with different QFs and embedded with different lengths of data using our scheme and four previous schemes.'-' means the capacity is not enough for the payload}
\label{tab:filesize}
\resizebox{\textwidth}{!}{
\begin{tabular}{@{}ccrrrrrrrrrrrrrrrr@{}}
\toprule
\multirow{2}{*}{\textbf{Image}} & \multirow{2}{*}{\textbf{Scheme}} & \multicolumn{4}{l}{\textbf{Payload(bits) with QF=30}}           & \multicolumn{4}{l}{\textbf{Payload(bits) with QF=50}}           & \multicolumn{4}{l}{\textbf{Payload(bits) with QF=70}}            & \multicolumn{4}{l}{\textbf{Payload(bits) with QF=90}}             \\ \cmidrule(l){3-18} 
                                &                                  & 3000           & 6000          & 9000          & 12000          & 4000          & 8000          & 12000          & 16000          & 5000          & 10000          & 15000          & 20000          & 6000           & 12000          & 18000          & 24000          \\ \midrule
Baboon                          & Huang $et~al.$\cite{IEEEexample:huang2016reversible}                            & 4200           & 8352          & 12920         & 17024          & 5712          & 10992         & 16904          & 23992          & 7408          & 15168          & 23312          & 31920          & 10264          & 20264          & 30832          & 41256          \\
                                & Qian $et~al.$\cite{IEEEexample:qian2017reversible}                             & 2992           & 7296          & 11624         & 14880          & 5040          & 9104          & 16568          & 20920          & 6824          & 13776          & 22808          & 28304          & 8456           & 20088          & 28512          & 37408          \\
                                & Wedaj $et~al.$\cite{IEEEexample:wedaj2017improved}                            & 3400           & 7560          & 12096         & 15760          & 5280          & 10688         & 16680          & 22800          & 8512          & 16504          & 24816          & 33640          & 14904          & 27624          & 38104          & 47176          \\
                                & Hou $et~al.$\cite{IEEEexample:hou2018reversible}                              & 3840           & 8312          & 12432         & 16416          & 5344          & 10680         & 16400          & 22936          & 6784          & 14464          & 22504          & 31224          & 9760           & 19600          & 30968          & 40776          \\
                                & Proposed (GFI,$m$=5)                        & 4063           & {\ul 7032}    & {\ul 11223}   & 14903          & 5959          & 11208         & 17664          & 23355          & 8307          & {\ul 13585}    & 23028          & 29780          & {\ul 6885}     & {\ul 17576}    & {\ul 26576}    & {\ul 32435}    \\
                                & Proposed (PFI,$m$=5)                        & \textbf{-3287} & \textbf{-318} & \textbf{3873} & \textbf{7553}  & \textbf{1610} & \textbf{6859} & \textbf{13315} & \textbf{19006} & \textbf{5989} & \textbf{11267} & \textbf{20710} & \textbf{27462} & \textbf{-2285} & \textbf{8406}  & \textbf{17406} & \textbf{23265} \\ \midrule
Boat                            & Huang $et~al.$\cite{IEEEexample:huang2016reversible}                            & 4384           & 8664          & 12496         & 16752          & 5832          & 11720         & 17848          & 24192          & 8096          & 15512          & 23272          & 31472          & 9496           & 20096          & 30944          & 40992          \\
                                & Qian $et~al.$\cite{IEEEexample:qian2017reversible}                             & 3680           & 8088          & 10488         & 14784          & 5040          & 10008         & 15304          & 23048          & \textbf{6176} & \textbf{12408} & \textbf{20448} & \textbf{29440} & 9152           & 18824          & 29728          & 37136          \\
                                & Wedaj $et~al.$\cite{IEEEexample:wedaj2017improved}                            & 3880           & 8224          & 11352         & 15968          & 5168          & 10904         & 16704          & 23592          & 6992          & 14592          & 22208          & 30568          & 9544           & 19824          & 28352          & 38952          \\
                                & Hou $et~al.$\cite{IEEEexample:hou2018reversible}                              & 3920           & 7784          & 12168         & 16192          & 5408          & 10920         & 16840          & 23848          & 7544          & 14608          & 22336          & 30728          & 8944           & 19680          & 30584          & 41024          \\
                                & Proposed (GFI,$m$=5)                        & 4018           & 8161          & 11548         & 16431          & 5476          & 10750         & 17061          & {\ul 22715}    & 8176          & 14681          & 23883          & 30685          & 11267          & 21661          & 31800          & 39753          \\
                                & Proposed (PFI,$m$=5)                        & \textbf{-365}  & \textbf{3778} & \textbf{7165} & \textbf{12048} & \textbf{3584} & \textbf{8858} & \textbf{15169} & \textbf{20823} & 7174          & 13679          & 22881          & 29683          & \textbf{6531}  & \textbf{16925} & \textbf{27064} & \textbf{35017} \\ \midrule
Elaine                          & Huang $et~al.$\cite{IEEEexample:huang2016reversible}                            & 4144           & 8288          & 12112         & ‒              & 6688          & 13024         & 18216          & 24952          & 8744          & 16200          & 23680          & 30000          & 9440           & 19160          & 28792          & 38680          \\
                                & Qian $et~al.$\cite{IEEEexample:qian2017reversible}                             & 3624           & 6608          & 12112         & ‒              & 5176          & 10992         & 15936          & 23608          & 6144          & 12416          & 18632          & 27544          & 7840           & 17736          & 27624          & 36952          \\
                                & Wedaj $et~al.$\cite{IEEEexample:wedaj2017improved}                            & 4376           & 8840          & 11696         & ‒              & 5568          & 11200         & 15928          & 23408          & 6024          & 11600          & 17576          & 25872          & 8416           & 17984          & 26416          & 34136          \\
                                & Hou $et~al.$\cite{IEEEexample:hou2018reversible}                              & 4720           & 8328          & 12160         & ‒              & 6120          & 12408         & 17944          & 24056          & 7536          & 14680          & 21672          & 28232          & 9088           & 18592          & 28496          & 38088          \\
                                & Proposed (GFI,$m$=5)                        & {\ul 3384}     & 6765          & {\ul 11265}   & {\ul 16372}    & 5357          & {\ul 10182}   & {\ul 15467}    & 24076          & 7019          & 14233          & 22491          & 27450          & 10020          & 21566          & 31678          & 41090          \\
                                & Proposed (PFI,$m$=5)                        & \textbf{-2509} & \textbf{872}  & \textbf{5372} & \textbf{10479} & \textbf{514}  & \textbf{5339} & \textbf{10624} & \textbf{19233} & \textbf{1970} & \textbf{9184}  & \textbf{17442} & \textbf{22401} & \textbf{2524}  & \textbf{14070} & \textbf{24182} & \textbf{33594} \\ \midrule
Lena                            & Huang $et~al.$\cite{IEEEexample:huang2016reversible}                            & 3784           & 8000          & 12568         & ‒              & 5568          & 11616         & 17680          & ‒              & 7080          & 14392          & 21648          & 30688          & 9336           & 17560          & 25464          & 35256          \\
                                & Qian $et~al.$\cite{IEEEexample:qian2017reversible}                             & 3528           & 8136          & 11952         & ‒              & 5232          & 10304         & 16832          & ‒              & 6952          & \textbf{12560} & \textbf{19952} & 30352          & 7136           & \textbf{14136} & \textbf{22936} & 34896          \\
                                & Wedaj $et~al.$\cite{IEEEexample:wedaj2017improved}                            & 4128           & 7672          & 12392         & ‒              & 5464          & 10808         & 17144          & ‒              & \textbf{6064} & 13712          & 21160          & 30192          & \textbf{6800}  & 15520          & 24160          & \textbf{33352} \\
                                & Hou $et~al.$\cite{IEEEexample:hou2018reversible}                              & 4056           & 7936          & 12624         & ‒              & 5488          & 11240         & 17232          & ‒              & 7440          & 14128          & 21368          & 30784          & 7624           & 16248          & 25192          & 34584          \\
                                & Proposed (GFI,$m$=5)                        & 3282           & 7551          & 13223         & {\ul 16942}    & 4712          & 10422         & 17505          & 22843          & 7176          & 14757          & 21666          & 31066          & 9114           & 18181          & 30641          & 39488          \\
                                & Proposed (PFI,$m$=5)                        & \textbf{-1714} & \textbf{2555} & \textbf{8227} & \textbf{11946} & \textbf{2045} & \textbf{7755} & \textbf{14838} & \textbf{20176} & 6279          & 13860          & 20769          & \textbf{30169} & 8384           & 17451          & 29911          & 38758          \\ \midrule
Peppers                         & Huang $et~al.$\cite{IEEEexample:huang2016reversible}                            & 3640           & 8232          & 12576         & ‒              & 5792          & 11104         & 17472          & ‒              & 7328          & 14440          & 21728          & 29432          & 9128           & 17808          & 26352          & 36744          \\
                                & Qian $et~al.$\cite{IEEEexample:qian2017reversible}                             & 3088           & 7264          & 12288         & ‒              & 4536          & 10992         & 16256          & ‒              & \textbf{5280} & \textbf{11664} & 20768          & 29536          & 7208           & 14080          & 23824          & 35160          \\
                                & Wedaj $et~al.$\cite{IEEEexample:wedaj2017improved}                            & 4712           & 8168          & 12312         & ‒              & 6160          & 10648         & 16696          & ‒              & 6376          & 12848          & \textbf{20352} & \textbf{28880} & 8352           & 16256          & 22864          & 31992          \\
                                & Hou $et~al.$\cite{IEEEexample:hou2018reversible}                              & 3520           & 7640          & 12568         & ‒              & 5808          & 10832         & 16712          & ‒              & 7104          & 13656          & 21184          & 28992          & 8288           & 16624          & \textbf{26192} & 36744          \\
                                & Proposed (GFI,$m$=5)                        & 3135           & 7344          & 13113         & {\ul 16783}    & 4779          & {\ul 10409}   & 17513          & 22984          & 6917          & 13011          & 21949          & 31873          & 9868           & 19569          & 32914          & 32914          \\
                                & Proposed (PFI,$m$=5)                        & \textbf{-1677} & \textbf{2532} & \textbf{8301} & \textbf{11971} & \textbf{2230} & \textbf{7860} & \textbf{14964} & \textbf{20435} & 5961          & 12055          & 20993          & 30917          & \textbf{3435}  & \textbf{13136} & \textbf{26481} & \textbf{26481} \\ \midrule
Tiffany                         & Huang $et~al.$\cite{IEEEexample:huang2016reversible}                            & 3760           & 8536          & 12976         & ‒              & 5640          & 12144         & 18608          & ‒              & 6824          & 14816          & 21672          & 30600          & 8720           & 16224          & 25304          & 35400          \\
                                & Qian $et~al.$\cite{IEEEexample:qian2017reversible}                             & 3168           & 6840          & 12232         & ‒              & 5160          & 11168         & 17280          & ‒              & 6784          & 14656          & 20080          & 29768          & \textbf{6432}  & \textbf{15608} & \textbf{22824} & \textbf{34848} \\
                                & Wedaj $et~al.$\cite{IEEEexample:wedaj2017improved}                            & 4088           & 7920          & 12232         & ‒              & 5272          & 11368         & 17200          & ‒              & 6048          & 13856          & 20232          & 29864          & 6864           & 16232          & 23448          & 33896          \\
                                & Hou $et~al.$\cite{IEEEexample:hou2018reversible}                              & 3544           & 7576          & 12696         & ‒              & 5736          & 11216         & 18168          & ‒              & 7032          & 14096          & 21536          & 30448          & 7760           & 16080          & 24840          & 34960          \\
                                & Proposed (GFI,$m$=5)                        & 4211           & 7130          & {\ul 11806}   & {\ul 15179}    & {\ul 4702}    & {\ul 10210}   & {\ul 16737}    & {\ul 21123}    & 6590          & {\ul 13809}    & 21992          & 30480          & 9119           & 22815          & 29846          & 41118          \\
                                & Proposed (PFI,$m$=5)                        & \textbf{-2797} & \textbf{122}  & \textbf{4798} & \textbf{8171}  & \textbf{-232} & \textbf{5276} & \textbf{11803} & \textbf{16189} & \textbf{3487} & \textbf{10706} & \textbf{18889} & \textbf{27377} & 8668           & 22364          & 29395          & 40667          \\ \bottomrule
\end{tabular}
}
\end{table*}

\par The file size increments of our proposed scheme and the four related DCT-based schemes are provided in Table \uppercase\expandafter{\romannumeral3}. It can be seen from Table \uppercase\expandafter{\romannumeral3} that the PFIs of most of the marked JPEG images generated by our scheme are much smaller than previous schemes. The lowest file size increments for different test images with different QFs are displayed in bold. Some file size increments are even negative, which means that the coding redundancy is larger than the file size increments caused by shifting and embedding. Additionally, the GFIs obtained by our scheme are close to the previous method, even smaller, which are displayed in underlined.
\par To further compare the performance of file size preservation, 200 images from the UCID image database are embedded into different payloads with different QFs. The average file size increments are shown in . It is observed from that the average PFIs obtained by the proposed method are smaller than the previous schemes at different payloads and QFs, which demonstrates that our proposed scheme is highly effective for file size preservation. 


\par The comparison of the performance of embedding capacity is also considered. We compare the embedding capacity with previous DCT-based schemes for the image Baboon at different QFs, which is illustrated in . It can be observed from  that the upper bound of embedding capacity of the previous DCT-based schemes is limited, while the FIs are also higher than the proposed scheme. Moreover, the upper bound of the embedding capacity of the proposed scheme can be higher if need. Note that there exists a sharp increment at the high payload for our proposed scheme, which means the performance of the proposed scheme after this sharp increment may not be an acceptable level. For the image Baboon with QF of 70, the PFI is close to GFI, which is because the coding redundancy is quite small.


\section{Conclusion}
\par In this paper, we proposed a high capacity lossless data hiding scheme for JPEG images by constructing the GVM relationship. Different from the previous arts, we allow the length of VLCs in one mapping set can be unequal. Therefore, the proposed scheme achieves the goal of high embedding capacity. Moreover, the corresponding RSVs before and after replacing are unchanged, the visual quality of the marked JPEG image can keep unchanged. The experimental results demonstrated that, at most of the time, the file size increments caused by the proposed scheme is smaller than the previous reversible data hiding schemes under the identical payload. Moreover, the upper bound of the embedding capacity of our proposed scheme is also much larger than state-of-the-arts works.
\par However, the limitation of our VLC-based scheme cannot be ignored. Our proposed scheme is only implemented on the JPEG images compressed by using the default Huffman table. Therefore, the proposed scheme does not apply to the JPEG images using the optimized Huffman table. In future work, we will utilize a more effective algorithm to solve the optimization problem of the proposed scheme more fastly. Besides, we will apply the proposed scheme to the JPEG images compressed by using the optimized Huffman table.

\section*{Acknowledgment}
\par This research work is partly supported by National Natural Science Foundation of China (61872003, U1636206).


\bibliographystyle{IEEEtran}
\bibliography{paper}

\end{document}